\documentclass[aps, 10pt, prl,
               notitlepage, twocolumn,
               superscriptaddress, longbibliography,
               floatfix, nobibnotes]{revtex4-2}
\usepackage{float}
\usepackage{placeins}
\usepackage{amsmath}
\usepackage{amssymb}
\usepackage{amsfonts}
\usepackage[utf8]{inputenc}
\usepackage[T1]{fontenc}
\usepackage{mathrsfs}
\usepackage{anyfontsize}
\usepackage{xspace}
\usepackage[inline]{enumitem}

\usepackage[linktocpage,breaklinks]{hyperref}
\usepackage[usenames,dvipsnames]{xcolor}

\usepackage{tensor}
\usepackage{bm}
\usepackage{dsfont}

\usepackage{graphicx}
\usepackage{epsfig}
\usepackage{epstopdf}

\hypersetup{colorlinks=true,
            citecolor=NavyBlue,
            linkcolor=NavyBlue,
            urlcolor=NavyBlue}

\begin{document}

\title{Dynamical scalarization in Schwarzschild binary inspirals}

\author{F\'elix-Louis Juli\'e}
\email{felix-louis.julie@aei.mpg.de}
\affiliation{Max Planck Institute for Gravitational Physics (Albert Einstein Institute),
\\ Am M\"uhlenberg 1, 14476 Potsdam, Germany}

\date{\today}

\begin{abstract}
We show that Schwarzschild black hole binaries can undergo dynamical scalarization (DS) in the inspiral phase, in a subclass of $\mathbb{Z}_2$-symmetric Einstein-scalar-Gauss-Bonnet (ESGB) theories of gravity.
The mechanism is analogous to neutron star DS in scalar-tensor gravity, and it differs from the late merger and ringdown black hole (de)scalarization found in recent ESGB studies.
To our knowledge, the new parameter space
we highlight was unexplored in numerical relativity simulations.
We also estimate the orbital separation at the DS onset, and characterize the subsequent scalar hair growth at the adiabatic approximation.
\end{abstract}

\maketitle

\noindent\textit{Introduction.}
Gravitational wave astronomy is a unique opportunity to probe gravity in the strong field of a compact binary coalescence~\cite{LIGOScientific:2016aoc,TheLIGOScientific:2017qsa,LIGOScientific:2020aai,LIGOScientific:2021qlt,LIGOScientific:2021djp,KAGRA:2023pio}.
While most tests so far constrain theory-agnostic deviations from general relativity (GR)~\cite{TheLIGOScientific:2016src,LIGOScientific:2019fpa,Mehta:2022pcn},
current efforts aim at calculating waveform templates in specific modified gravity theories.

Scalar-tensor gravity (ST) is perhaps the simplest and most studied example~\cite{Goenner:2012cq,Will:2014kxa,Damour:1992we}.
The two-body dynamics has indeed been addressed within
the post-Newtonian (PN) and effective-one-body (EOB) frameworks~\cite{Damour:1992we,Damour:1995kt,Mirshekari:2013vb,Lang:2013fna,Lang:2014osa,Sennett:2016klh,Julie:2017pkb,Julie:2017ucp,Bernard:2018hta,Bernard:2018ivi,Bernard:2019yfz,Bernard:2022noq,Jain:2022nxs,Julie:2022qux,Jain:2023fvt,Jain:2023vlf,Bernard:2023eul}.
Interestingly, isolated neutron stars (NSs) can undergo spontaneous scalarization, i.e. grow nonzero scalar hair due to a symmetry breaking~\cite{Damour:1993hw}.
Another abrupt mechanism, named dynamical scalarization (DS), was found in numerical relativity (NR) simulations of NS binaries~\cite{Barausse:2012da,Shibata:2013pra,Palenzuela:2013hsa} in models escaping solar system and binary pulsar tests~\cite{Bertotti:2003rm,Esposito-Farese:2009ouh,Freire:2012mg}.
When DS happens, the inspiraling system is first the same as in GR, until a nontrivial scalar field configuration becomes energetically favorable at some orbital separation.
Various methods were proposed to include DS in waveform models~\cite{Sennett:2016rwa,Sennett:2017lcx,Khalil:2019wyy,Khalil:2022sii}.

Yet, in ST gravity, static black holes (BHs) are the same as in GR.
By contrast, Einstein-scalar-Gauss-Bonnet (ESGB) theories allow for hairy BHs~\cite{Mignemi:1992nt,Torii:1996yi,Yunes:2011we,Sotiriou:2013qea,Sotiriou:2014pfa,Kanti:1995vq,Pani:2009wy,Pani:2011gy,Ayzenberg:2014aka,Maselli:2015tta,Kleihaus:2015aje,Antoniou:2017acq,Cunha:2019dwb,Julie:2019sab,Julie:2022huo}, and certain subclasses predict the spontaneous scalarization of the Schwarzschild or Kerr spacetimes~\cite{Doneva:2017bvd,Silva:2017uqg,Minamitsuji:2018xde,Silva:2018qhn,Macedo:2019sem,Minamitsuji:2019iwp,Dima:2020yac}.
Recent progress was made in the PN and EOB contexts~\cite{Julie:2019sab,Julie:2022qux,Shiralilou:2021mfl, Julie:2022huo,vanGemeren:2023rhh} and in NR~\cite{Okounkova:2017yby,Witek:2018dmd,Okounkova:2019dfo,Julie:2020vov,Witek:2020uzz,Okounkova:2020rqw,East:2020hgw,East:2021bqk,Figueras:2021abd,Corman:2022xqg,AresteSalo:2022hua,Brady:2023dgu,Doneva:2023oww,AresteSalo:2023mmd}.
New effects were pointed out numerically, such as spin-induced dynamical scalarization, descalarization, and stealth scalarization, but they were only observed in the late merger or ringdown phase of a BH coalescence~\cite{Silva:2020omi,Doneva:2022byd,Elley:2022ept,Doneva:2022ewd}.

Here, we report a different mechanism: the \textit{DS of Schwarzschild BH binaries}, which resembles that of NSs in ST theories, and takes place in the inspiral phase.
In order to study DS analytically, we shall first
derive the \textit{sensitivities} of a Schwarzschild BH.
Throughout
this Letter we use geometrical units ($G=c=1$).\\

\noindent\textit{Nonrotating Einstein-scalar-Gauss-Bonnet black holes.}
We consider the theory described by the action
\begin{align}
I  = \frac{1}{16\pi}  \int\! d^{4}x & \sqrt{-g} \big(R - 2g^{\mu\nu} \partial_{\mu}\varphi \partial_{\nu}\varphi +\ell^2 f(\varphi)\mathcal G\big)\,,
\label{eq:action}
\end{align}
where
$
\mathcal G=R^{\mu\nu\rho\sigma}R_{\mu\nu\rho\sigma} - 4R^{\mu\nu}R_{\mu\nu} + R^2
$ 
is the Gauss-Bonnet scalar, whose integral $\int d^4x\sqrt{-g}\,\mathcal G$ 
is a boundary term \cite{Myers:1987yn}.
The dimensionless function $f$ (defined modulo a constant) and the length $\ell$ specify the theory.
The field equations obtained by varying~\eqref{eq:action} are:
\begin{subequations}
\begin{align}
R_{\mu\nu}&=2\partial_\mu\varphi\partial_\nu\varphi-4\ell^2(P_{\mu\alpha\nu\beta}-\tfrac{1}{2}g_{\mu\nu}P_{\alpha\beta})\nabla^\alpha\nabla^\beta f\,,\label{eq:EinsteinEquations}\\
    \Box\varphi&=-\tfrac{1}{4}\ell^2 f_{,\varphi}(\varphi)\,\mathcal G\,,\label{eq:KGEquation}
\end{align}\label{eq:fieldEquations}%
\end{subequations}
where the tensor $P^{\mu}{}_{\nu\rho\sigma}$ with trace $P_{\mu\nu}$ was introduced in~\cite{Davis:2002gn}, and where we denote $(\cdot)_{,\varphi}=d(\cdot)/d\varphi$.

We focus on asymptotically flat, static, spherically symmetric BHs.
We consider the line element in Schwarzschild-Droste coordinates
\begin{align}
    ds^2=-N(r)\sigma^2(r)dt^2+N^{-1}(r)dr^2+r^2 d\Omega^2\,,\label{eq:metricAnsatz}
\end{align}
and scalar field $\varphi=\varphi(r)$.
A few reminders are in order.
The solutions to Eqs.~\eqref{eq:fieldEquations} such that $N$ features a simple zero at the horizon radius $r=r_{\rm H}$ depend on two integration constants~\cite{Julie:2019sab,Julie:2022huo}.
The first law of thermodynamics of these BHs is found as follows:
Define the entropy à la Wald~\cite{Wald:1993nt,Iyer:1994ys,Torii:1996yi}
\begin{align}
    \mathscr S_{\rm w}=\frac{1}{4}\mathcal A_{\rm H}+4\pi\ell^2 f(\varphi_{\rm H})\,,\label{eq:WaldEntropy}
\end{align}
where $\mathcal A_{\rm H}$ is the horizon area, and where $\varphi_{\rm H}=\varphi(r_{\rm H})$.
Define also the Arnowitt-Deser-Misner (ADM) mass $m$, scalar charge $d$ and scalar background $\bar\varphi$ from the asymptotic behavior of $N$ and $\varphi$ at spatial infinity:
\begin{align}
N=1-\frac{2m}{r}+\mathcal O(r^{-2})\,,\quad 
\varphi=\bar\varphi+\frac{d}{r}+\mathcal O(r^{-2})\,.\label{eq:NandBarVarphi}
\end{align}%
Ref.~\cite{Julie:2019sab} then showed that the variation of $\mathscr S_{\rm w}$, $m$ and $\bar\varphi$ with respect to the BH's two integration constants satisfies the identity
\begin{align}
    T\delta  \mathscr S_{\rm w}=\delta m+d\,\delta\bar\varphi\,,\label{eq:firstLaw}
\end{align}
where $T$ is the temperature, whose expression we do not need here.

Nonrotating BHs were obtained analytically in the small-$\ell$ limit for $f(\varphi)=2\varphi$ \cite{Yunes:2011we,Sotiriou:2013qea,Sotiriou:2014pfa}, for $f(\varphi)=\frac{1}{4}\exp(2\varphi)$ \cite{Mignemi:1992nt,Torii:1996yi} and for a generic $f(\varphi)$~\cite{Julie:2019sab}.
They were also extended numerically to the nonpertubative regime~\cite{Kanti:1995vq,Pani:2009wy,Antoniou:2017acq,Julie:2022huo}.
In particular, theories with $f_{,\varphi}(0)=0$
and $f_{,\varphi\varphi}(0)\mathcal G>0$
can predict the spontaneous scalarization of the Schwarzschild spacetime.
This is the case with
$f(\varphi)=\frac{1}{2\lambda}[1-\exp(-\lambda\varphi^2)]$ and $f(\varphi)=\frac{1}{2}\varphi^2-\frac{\lambda}{4}\varphi^4$, where $\lambda>0$ ensures the stability of the resulting hairy configurations~\cite{Doneva:2017bvd,Silva:2017uqg,Minamitsuji:2018xde,Silva:2018qhn,Macedo:2019sem,Minamitsuji:2019iwp,Julie:2022huo}.
However, the latter works are restricted to BHs with $\bar\varphi=0$ (an exception being~\cite{Julie:2022huo} in the Gaussian case with $\lambda=6$).
For the inclusion of spins, cf.~\cite{Ayzenberg:2014aka,Pani:2011gy,Maselli:2015tta,Kleihaus:2015aje,Cunha:2019dwb,Dima:2020yac}.\\

\noindent\textit{The sensitivities of a Schwarzschild black hole.}
In this Letter, we examine the subclass of theories whose action \eqref{eq:action} is invariant under the $\mathbb{Z}_2$ transformation $\varphi\to -\varphi$, with $f_{,\varphi\varphi}(0)>0$, and focus on their small-$\varphi$ limit.
Through quartic order we
can write, without loss of generality:
\begin{align}
f(\varphi)=\frac{1}{2}\varphi^2-\frac{\lambda}{4}\varphi^4+\cdots\,.\label{eq:Z2theory}%
\end{align}
Our results will thus apply to
the theories studied in Refs.~\cite{Silva:2017uqg,Doneva:2017bvd,Minamitsuji:2018xde,Silva:2018qhn,Macedo:2019sem,Minamitsuji:2019iwp,Silva:2020omi,Doneva:2022byd,Julie:2022huo,East:2021bqk}.
The Schwarzschild spacetime is a particular solution of the subclass when $\varphi=0$, since then~\eqref{eq:EinsteinEquations} reduces to $R_{\mu\nu}=0$ and $f_{,\varphi}(\varphi)$ vanishes in~\eqref{eq:KGEquation}.
Yet, we must 
address the following issue, as a prerequisite to study dynamical scalarization:
How does such Schwarzschild BH respond to a slow, adiabatic increase in its
(initially zero)
scalar background $\bar\varphi$\,?

To answer this question,
we derive here a sequence of BH solutions with fixed Wald entropy $\mathscr S_{\rm w}$, but \textit{generically nonzero} $\bar\varphi$.
These are our two BH integration constants.
A suitable dimensionless quantity to describe the sequence
is~\cite{Julie:2019sab,Cardenas:2017chu,Julie:2022huo}
\begin{align}
    \alpha
    =\frac{1}{m}\left.\frac{\partial m}{\partial\bar\varphi} \right\vert_{\mathscr S_{\rm w}}\,,
    \label{eq:sensitivityDef}%
\end{align}
which simplifies as $\alpha=-(d/m)$ as a consequence of the first law of thermodynamics~\eqref{eq:firstLaw} when $\delta \mathscr S_{\rm w}=0$.

Given the expansion~\eqref{eq:Z2theory},
%and below,
we can derive
$\alpha$
for $\bar\varphi\ll 1$, around the Schwarzschild spacetime recovered at
$\bar\varphi=0$.
Using the $\mathbb{Z}_2$ symmetry of the theory, we anticipate that $\alpha\to -\alpha$ when $\bar\varphi\to -\bar\varphi$.
Thus we must have:
\begin{align}
\alpha=\beta_{(1)}\bar\varphi+\frac{1}{6}\beta_{(3)}\bar\varphi^3+\cdots\,.\label{eq:alphaExpansion}
\end{align}
The coefficients $\beta_{(1)}$ and $\beta_{(3)}$ depend on
%the constant Wald entropy
$\mathscr S_{\rm w}$ (along with the theory parameters
$\ell$ and $\lambda$), and we shall refer to them as the \textit{first and second sensitivities} of a Schwarzschild BH.
They characterize its adiabatic response to a small but nonzero $\bar\varphi$ through cubic order.

From now on, a subscript $0$ indicates a quantity evaluated at $\bar\varphi=0$, as the BH reduces to Schwarzschild.
We also conveniently trade the fixed $\mathscr S_{\rm w}$ labeling our sequence for the value $m_0$ of the ADM mass in the Schwarzschild limit.
Indeed, they are
related through~\eqref{eq:WaldEntropy} by $m_0=(\mathscr S_{\rm w}/4\pi)^{1/2}$.
We recognize the irreducible mass in the right member~\cite{Christodoulou:1970wf}.
We calculate $\beta_{(1)}$ and $\beta_{(3)}$ in two independent ways:
\begin{enumerate}
\item We follow the analytical methods detailed in Ref.~\cite{Julie:2019sab} to calculate $\alpha$ in the subclass~\eqref{eq:Z2theory}, for all $\lambda$ and in a small-$(\ell/m_0)^2$ expansion.
Then, by identifying the result with~\eqref{eq:alphaExpansion}, the sensitivities read
\begin{align}
    \beta_{(i)}=\sum_{n=1}^{N} b_n^{(i)} (\ell/m_0)^{2n}\,, \label{eq:sensiSeries}%
\end{align}
which we push here to $N=14$.
The $b_n^{(1)}$ are rational numbers, while only the $b_n^{(3)}$ depend on $\lambda$.
We provide their lengthy expressions online~\cite{FLJRepoDS}.
We finally accelerate the convergence of the series~\eqref{eq:sensiSeries} for large $(\ell/m_0)^2$ values using the $(7,7)$-Pad\'e approximants $\beta_{(i)}^{\rm Pad\acute{e}}$ discussed in
the Supplemental Material.
\item We apply the numerical methods of Ref.~\cite{Julie:2022huo} to derive BH sequences
with fixed Wald entropy in the subclass~\eqref{eq:Z2theory}.
We focus on the cases $\lambda=4,6,8$, where $\lambda=6$ corresponds the Gaussian theory studied in Refs.~\cite{Doneva:2017bvd,Julie:2022huo,Doneva:2022byd}.
We skim through $(\ell/m_0)^2$ values
with increment $\Delta (\ell/m_0)^2=0.02$.
For each value, we derive $\alpha$ on a small-$\bar\varphi$ range, and calculate $\beta_{(1)}^{\rm Num}=(\partial\alpha/\partial\bar\varphi)_0$ and $\beta_{(3)}^{\rm Num}=(\partial^3\alpha/\partial\bar\varphi^3)_0$, see the Supplemental Material.
\end{enumerate}

\begin{figure*}[ht]
\includegraphics[width=1.0\columnwidth]{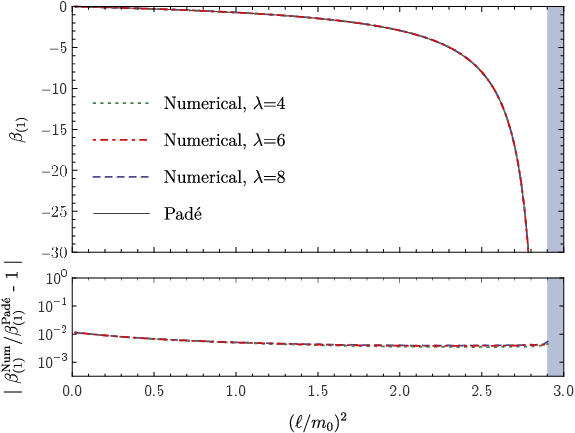}
\quad
\includegraphics[width=1.0\columnwidth]{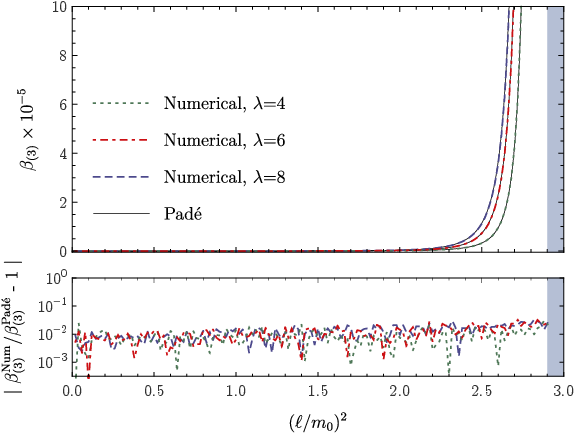}
\caption{Upper panels: First and second sensitivity of a Schwarzschild BH with ADM mass $m_0$.
They vanish for $(\ell/m_0)^2=0$, and their $(7,7)$-Pad\'e approximants are singular at the threshold~\eqref{eq:spontScalThreshold}.
We discard larger $(\ell/m_0)^2$ values, since then spontaneous scalarization occurs (dark-shaded).
Lower panels: the fractional errors between the numerical results $\beta_{(i)}^{\rm Num}$ and their analytical counterparts $\beta_{(i)}^{\rm Pad\acute{e}}$ reach at most $3\%$.
}
\label{fig:betas}%
\end{figure*}%

The results are displayed in Fig.~\ref{fig:betas}.
The numerical calculations show remarkable agreement with the $(7,7)$-Pad\'e approximants.
The coupling $\ell$ and Schwarzschild mass $m_0$ only contribute through the ratio $(\ell/m_0)^2$.
The left panel evidences that $\beta_{(1)}^{\rm Num}$ is not affected by $\lambda$, consistently with our analytical results (cf.~\eqref{eq:sensiSeries} and below).
The reason is that $\beta_{(1)}$ can be calculated by linearizing Eqs.~\eqref{eq:fieldEquations} with respect to $\varphi$.
This amounts to solving \eqref{eq:KGEquation} in the Schwarzschild metric background with~\eqref{eq:Z2theory} truncated to $\mathcal O(\varphi^2)$.
Likewise,
we checked analytically that
$\beta_{(3)}$ depends on $\ell$ and $\lambda$, but is unchanged upon adding $\mathcal O(\varphi^{2n})$ corrections with $n\ge 3$ in~\eqref{eq:Z2theory}.
Our sensitivities are thus \textit{universal} to the entire subclass of $\mathbb{Z}_2$-symmetric theories considered here.

The sensitivities vanish in the ST limit $(\ell/m_0)^2=0$, and most remarkably, their $(7,7)$-Pad\'e approximants predict a pole, of order one and four respectively, at
\begin{align}
    (\ell/m_0)_{\rm pole}^2\simeq 2.90\,.\label{eq:spontScalThreshold}%
\end{align}
We recover here, with excellent agreement, the spontaneous scalarization threshold
of Refs.~\cite{Doneva:2017bvd,Silva:2017uqg,Silva:2018qhn,Macedo:2019sem}.
When this threshold is exceeded, the Schwarzschild spacetime is unstable, and BH solutions with scalar hair yet $\bar\varphi=0$ appear.
Below, we will focus on BH binaries which, at least in the early inspiral phase, reduce to GR.
We thus restrict $(\ell/m_0)^2$ to values below the threshold \eqref{eq:spontScalThreshold} for each BH, so that they are Schwarzschild at infinite separation.
The discarded range is dark-shaded in Fig.~\ref{fig:betas}.
Finally, observe that $\beta_{(1)}\le 0$, and that $\beta_{(3)}\ge 0$ for $\lambda=4,6,8$.
From now on, we focus on these illustrative values, but our analytical results $\forall\lambda$ are available online~\cite{FLJRepoDS}.\\

\noindent\textit{Dynamical scalarization.}
The dynamics of inspiraling ESGB BH binaries was studied analytically within the PN framework~\cite{Yagi:2011xp,Julie:2019sab,Shiralilou:2020gah,Shiralilou:2021mfl,Julie:2022qux,vanGemeren:2023rhh} and BH perturbation theory~\cite{Maselli:2020zgv,Barsanti:2022ana}.
In this context, Refs.~\cite{Julie:2019sab,Cardenas:2017chu} showed that when neglecting tidal and out-of-equilibrium corrections, each BH is described by a sequence of equilibrium configurations
with fixed Wald entropy.

Consider now two inspiraling BHs, say $A$ and $B$, in the subclass~\eqref{eq:Z2theory}.
Given the above, the system is characterized by two copies of~\eqref{eq:alphaExpansion}:
\begin{subequations}
\begin{align}
\alpha_A&=\beta^A_{(1)}\bar\varphi_A+\frac{1}{6}\beta^A_{(3)}\bar\varphi_A^3+\cdots\,,\label{eq:sensiA}\\
\alpha_B&=\beta^B_{(1)}\bar\varphi_B+\frac{1}{6}\beta^B_{(3)}\bar\varphi_B^3+\cdots\,.\label{eq:sensiB}%
\end{align}
\label{eq:sensiAandB}%
\end{subequations}
From \eqref{eq:sensitivityDef},  we also have that
the ADM masses are $m_A=m_A^0\big[1+\frac{1}{2}\beta^A_{(1)}\bar\varphi_A^2+\cdots\big]$ and its $B$-counterpart.
The sensitivities are determined from Fig.~\ref{fig:betas}, once $\lambda$
and the ratios
$(\ell/m_{A}^0)^2$ and $(\ell/m_{B}^0)^2$ are chosen.
As for the scalar backgrounds $\bar \varphi_{A}$ and $\bar \varphi_{B}$ of each BH, they are now induced by the scalar hair of the respective companion~\cite{Julie:2019sab,Julie:2022huo}.
For simplicity, we truncate them to Coulomb level, and they read (cf. below~\eqref{eq:sensitivityDef}):

\begin{subequations}
\begin{align}
\bar\varphi_A&=\varphi(t,\mathbf x_A)\simeq-\frac{m_B\alpha_B}{r}\,,\\
\bar\varphi_B&=\varphi(t,\mathbf x_B)\simeq-\frac{m_A\alpha_A}{r}\,,\label{eq:CoulombBackgroundB}
\end{align}
\label{eq:CoulombBackgrounds}%
\end{subequations}
where $r=|\mathbf x_A-\mathbf x_B|$ is the orbital separation.

Let us introduce, in the Schwarzschild limit, the total mass $M=m_A^0+m_B^0$, mass ratios $q=m_A^0/m_B^0\ge 1$ and $\nu=m_A^0 m_B^0/M^2$, and the reduced orbital separation $\hat r=r/M$.
Inserting~\eqref{eq:sensiAandB} and below into~\eqref{eq:CoulombBackgrounds} yields a system of two equations for two unknowns $\bar\varphi_{A}$ and $\bar\varphi_{B}$,
which can be plugged into one another to give, e.g., the following equation on $\bar\varphi_A$:
\begin{align}
\bar\varphi_A\bigg[\bigg(\frac{\hat r^2}{\hat r_{\rm DS}^2}-1\bigg)
-\bar\varphi_A^2\,
\mathcal C(\hat r)+\cdots\bigg]=0\,,
\label{eq:phiA}%
\end{align}
where
\begin{align}
    \hat r_{\rm DS}&=\sqrt{\beta^{A}_{(1)}\beta^{B}_{(1)}\nu}\,,\label{eq:DSonset}
\end{align}
and where $\mathcal C(\hat r)$ is given in the Supplemental Material.
It is negative given the signs of the sensitivities in Fig.~\ref{fig:betas}.

As expected, the Schwarzschild configuration $\bar\varphi_A=0$ (and thus $\bar\varphi_B=0$ also, as seen by setting $\bar\varphi_A=0$ in~\eqref{eq:sensiA} and using \eqref{eq:CoulombBackgroundB}) is a solution for all $\hat r$.
However, when $\hat r<\hat r_{\rm DS}$, the situation changes:
Eq.~\eqref{eq:phiA} then features two additional nonzero, equal and opposite roots $\bar\varphi_A\neq0$ (and thus $\bar\varphi_B\neq 0$ also).
They describe energetically favorable
%hairy BH
equilibrium
configurations, since $\beta_{(1)}^{A}$ and $\beta_{(1)}^{B}$ entering the ADM masses below~\eqref{eq:sensiAandB} are negative \footnote{We neglect PN corrections to the system's energy, as done for ST theories in~\cite{Sennett:2017lcx}, cf. above Eq.~(32) and Eq.~(33) there.}.

\begin{figure*}[ht]
\includegraphics[width=0.9\columnwidth]{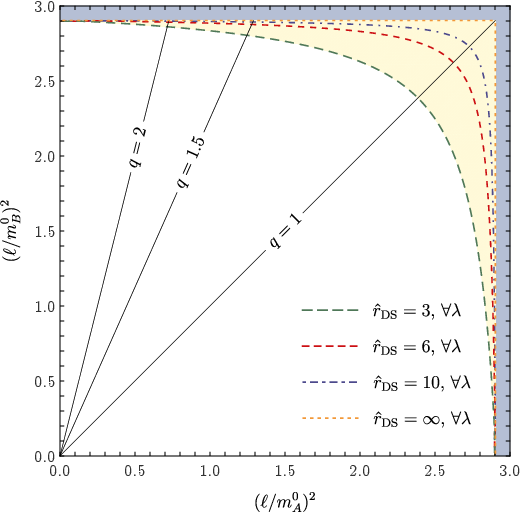}
\quad
\includegraphics[width=1\columnwidth]{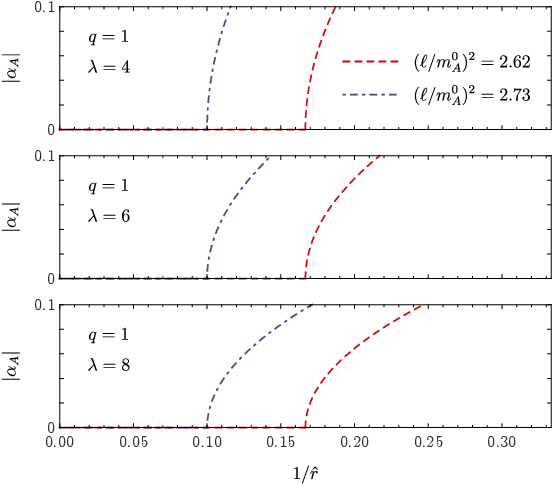}
\caption{Left panel:
Parameter space of two Schwarzschild BHs with ADM masses
$m_A^0$ and $m_B^0$.
In the light-shaded region,
%$\hat r_{\rm DS}\ge 3$ and
DS occurs in the inspiral phase.
We show some contour lines of $\hat r_{\rm DS}$, which diverges if either $(\ell/m_A^0)^2$ or $(\ell/m_B^0)^2$ approach
the
threshold 
\eqref{eq:spontScalThreshold}.
We discard larger values (dark-shaded region), as in Fig.~\ref{fig:betas}.
Right panels:
$|\alpha_A|$ against $1/\hat r$ for DS scenarios with $\hat r_{\rm DS}=6,10$.
}%
\label{fig:DsOnsetAndGrowth}%
\end{figure*}%

We refer to this mechanism as
the \textit{dynamical scalarization} (DS) of ESGB BHs, by analogy with the phenomenon discovered in Refs.~\cite{Barausse:2012da,Shibata:2013pra,Palenzuela:2013hsa,Taniguchi:2014fqa} for NSs in ST theories.
Note that~\eqref{eq:DSonset} is a function of $\nu$ and
the first sensitivities only, which do not depend on $\lambda$.

The left panel of 
Fig.~\ref{fig:DsOnsetAndGrowth} evaluates
$\hat r_{\rm DS}$ for a BH binary, given $(\ell/m_A^0)^2$ and $(\ell/m_B^0)^2$.
In the white region, $\hat r_{\rm DS}<3$, which we
choose to
discard.
Indeed, $\hat r_{\rm LR}=3$ is the light-ring radius in the EOB formalism at 1PN,
and it
marks the transition to the remnant's ringdown in a Schwarzschild binary evolution~\cite{Buonanno:1998gg,Buonanno:2000ef}.
By contrast, in the light-shaded region, $\hat r_{\rm DS}\ge 3$.
There, each point corresponds to a Schwarzschild binary that undergoes DS \textit{in the inspiral phase}.
The relevant range is largest when $q=1$, finding $2.39\lesssim(\ell/m_A^0)^2\lesssim 2.90$, and it shrinks as $q$ is increased.
We show the contour lines $\hat r_{\rm DS}=3, 6, 10$.
When $q=1$, the BHs are identical and \eqref{eq:DSonset} simplifies as $\hat r_{\rm DS}= -\beta_{(1)}^A/2$, which can be evaluated from the left panel of Fig.~\ref{fig:betas}.
More importantly, observe that
DS can take place arbitrarily early in the inspiral phase: indeed, $\hat r_{\rm DS}$ diverges when either $(\ell/m_A^0)^2$ or $(\ell/m_B^0)^2$ approach the threshold~\eqref{eq:spontScalThreshold}.
In the dark-shaded region, it is exceeded by at least one BH, meaning that the latter is already spontaneously scalarized at infinite separation. Thus we discard this region, as in Fig.~\ref{fig:betas}.

Knowing the second sensitivities allows us to characterize the scalar charge growth subsequent to the DS onset.
Solving for $\bar\varphi_A\neq 0$ in~\eqref{eq:phiA} and inserting the result into~\eqref{eq:sensiA} yields, to leading order in $0\le \hat r^{-1}-\hat r_{\rm DS}^{-1}\ll 1$:
\begin{align}
\alpha_A=\pm\, \Gamma_A \left(\frac{1}{\hat r}-\frac{1}{\hat r_{\rm DS}}\right)^{1/2}+\cdots\,,\label{eq:alphaDS}
\end{align}
where
\begin{equation}
\Gamma_A=\left[\frac{-2(\beta^A_{(1)})^2(\beta^{A}_{(1)}\beta^{B}_{(1)}\nu)^{1/2}}{
    \frac{1}{2}\beta^{A}_{(1)}+\frac{1}{6}\frac{\beta^A_{(3)}}{\beta^A_{(1)}}
    +q\Big(\frac{1}{2}\beta^{B}_{(1)}+\frac{1}{6}\frac{\beta^B_{(3)}}{\beta^B_{(1)}}\Big)\frac{\beta^A_{(1)}}{\beta^B_{(1)}}
}\right]^{1/2}\,,\label{eq:GammaA}
\end{equation}
and
$A\leftrightarrow B$ (i.e. $q\to q^{-1}$) by
elementary manipulations of~\eqref{eq:sensiAandB} and \eqref{eq:CoulombBackgrounds}.

The right panel of Fig.~\ref{fig:DsOnsetAndGrowth}
shows the magnitude of $\alpha_A$ against $1/\hat r$ in our DS scenario.
We take
identical BHs, $q=1$, such that~\eqref{eq:GammaA} simplifies as $\Gamma_A=\big[1/(\beta_{(1)}^A)^2+\frac{1}{3}\beta_{(3)}^A/(\beta_{(1)}^A)^4\big]^{-1/2}$.
We set $\lambda=4,6,8$, and choose $(\ell/m_A^0)^2$ values corresponding to $\hat r_{\rm DS}=6,10$, in the left panel.
The scalar charge growth is abrupt as $\hat r$ is decreased through $\hat r_{\rm DS}$, but it is mitigated by increasing $\lambda$.
Observe that the system exceeds $|\alpha_A|=0.1$ before the light-ring $\hat r_{\rm LR}=3$.
We do not display $|\alpha_A|\ge 0.1$ for consistency with our small-$\varphi$ assumption.

In the Supplemental Material, we evaluate $\Gamma_A$ and $\Gamma_B$ for various $q$ and $\lambda$ values, in the full relevant $(\ell/m_A^0)^2$ range.
We also discuss the effect of adding a homogeneous scalar background $\varphi_0$ on the system, and widen our results numerically, beyond the small-$\varphi$ regime.\\

\noindent\textit{Discussion.}
To our knowledge, we highlight here a DS mechanism \textit{in the inspiral phase} that was overlooked
in the literature on $\mathbb{Z}_2$-symmetric ESGB gravity.
Indeed, Refs.~\cite{Silva:2020omi,Doneva:2022byd,East:2021bqk} numerically simulate nonspinning BH
binaries for various initial data
\footnote{We translate our conventions into those of~\cite{Silva:2020omi,Doneva:2022byd,East:2021bqk} via $(\ell^2,\,\lambda)=(2\alpha_{\rm GB}\bar\beta_2,\,0)$, $(\bar\lambda^2,\,\beta-2\kappa)$, and $(4\bar\lambda,\,-2\sigma/\bar\lambda)$ respectively, adding a bar on their $\lambda$ to avoid confusing it with ours.}.
Yet, either $(\ell/m_A^0)^2$ or $(\ell/m_B^0)^2$ is taken above the threshold~\eqref{eq:spontScalThreshold}, or $\ell$ is set to zero, falling into the dark-shaded or white regions of Fig.~\ref{fig:DsOnsetAndGrowth}.
As for Refs.~\cite{Elley:2022ept,AresteSalo:2023mmd}, they study spinning BH binaries in theories with $f_{,\varphi\varphi}(0)<0$, which do not belong to the subclass~\eqref{eq:Z2theory}.
With such settings, these works identify \textit{late merger} and \textit{ringdown} mechanisms, named spin-induced dynamical scalarization, descalarization, and stealth scalarization.
They amount to considering binary systems made of BHs individually inside the spontaneous scalarization window (given their masses and spins), forming a remnant outside of it, or vice-versa.
We conclude that the light-shaded region in the left panel of Fig.~\ref{fig:DsOnsetAndGrowth} is new to this Letter.
It signals a different
mechanism
taking place in the inspiral,
that should be further explored in the future, both analytically and through NR simulations.

Importantly, we note that~\eqref{eq:DSonset} and \eqref{eq:alphaDS} 
hold for generic compact binary systems in all $\mathbb{Z}_2$-symmetric theories with a massless scalar field beyond ESGB.
Our work
provides a way to identify the DS parameter space, which is important to guide future NR developments in such theories.
A prerequisite is to calculate the sensitivities of a BH with fixed
(Wald) entropy, or of a NS with fixed baryonic number~\cite{Damour:1992we,Damour:1995kt,Damour:1996ke}.

Ref.~\cite{Sennett:2017lcx} studied NS binaries in ST gravity close to the DS onset, via an effective action modeling the scalar charges as extra degrees of freedom.
Later on, Ref.~\cite{Khalil:2019wyy} argued that theories predicting spontaneous scalarization can also exhibit DS, taking Reissner-Nordstr\"{o}m binaries in Einstein-Maxwell-scalar (EMS) gravity as an explicit illustration.
In a companion paper, we will show that
our sensitivities can be translated into the quantities $c_{(2)}$ and $c_{(4)}$ of these works.
We will also discuss further $\lambda$ values, and widen our results to the post-adiabatic approximation by generalizing to ESGB BHs the quantity $c_{\dot q^2}$ derived in~\cite{Khalil:2022sii} for ST NSs.
Note that the present work already differs from Refs.~\cite{Sennett:2017lcx,Khalil:2019wyy,Khalil:2022sii} in various ways:
(i) For the first time, we highlight explicitly and quantitatively the \textit{DS of Schwarzschild BHs}.
The latter do not carry constant electromagnetic or dark-sector $U(1)$ charges;
(ii) To do so, we crucially fix their \textit{Wald} entropies in ESGB theories, rather than Bekenstein;
(iii) Eqs.~\eqref{eq:DSonset}, \eqref{eq:alphaDS} and \eqref{eq:GammaA} are valid for \textit{asymmetric binaries} $A\neq B$, rather than only symmetric;
(iv) We calculate the BH sensitivities numerically and \textit{also analytically} for all $(\ell/m_0)^2$ and $\lambda$ values;
(v) We
explore extensively
the two-body parameter space, even when $q\neq 1$;
(vi) Our quantitative results apply to a \textit{full subclass}~\eqref{eq:Z2theory}, beyond specific theory examples;
(vii) We describe DS in the language of the PN sensitivities.

The effective-one-body (EOB) framework was recently extended to ESGB BH binaries in Ref.~\cite{Julie:2022qux}.
We leave to future work the inclusion of DS within the EOB formalism.
It will also be interesting to widen our analysis towards spinning BHs, and to the class of $\mathbb{Z}_2$-symmetric ESGB theories with $f_{,\varphi\varphi}(0)<0$.\\

\noindent\textit{Acknowledgements.}
I am grateful to Alessandra Buonanno, Maxence Corman, Mohammed Khalil, Hector O. Silva, and Jan Steinhoff for useful discussions.
This project is supported by the German Research Foundation (Deutsche
Forschungsgemeinschaft, DFG, Project No. 386119226).

\bibliography{FLJbib}

\pagebreak

\widetext

\vfill\eject

\section{Supplemental Material}

\subsection{Analytical sensitivities and Pad\'{e} approximants\label{App:Pade}}

To extend the scope of the series~\eqref{eq:sensiSeries} to nonperturbative $(\ell/m_0)^2$ values, we use diagonal Pad\'e approximants with respect to the variable $(\ell/m_0)^2$.
For $1\le k\le 7$ we find that $\mathcal P^k_k[\beta_{(1)}]$ features excellent convergence as $k$ is increased.
For example, the fractional error between the approximants $k=6$ and $k=7$ is smaller that $10^{-12}$ in the range $0<(\ell/m_0)^2<2.90$ relevant to this work.
In this Letter, we thus use:
\begin{align}
    \beta_{(1)}^{\rm Pad\acute{e}}&=\mathcal P^7_7[\beta_{(1)}]\,,\label{eq:beta1PadeDef}
\end{align}
which features a simple pole at the spontaneous scalarization threshold \eqref{eq:spontScalThreshold}.
Then~\eqref{eq:beta1PadeDef} exhibits striking agreement with our numerical results, cf. Fig.~\ref{fig:betas}.

\begin{figure*}[h!]
\includegraphics[width=0.48\columnwidth]{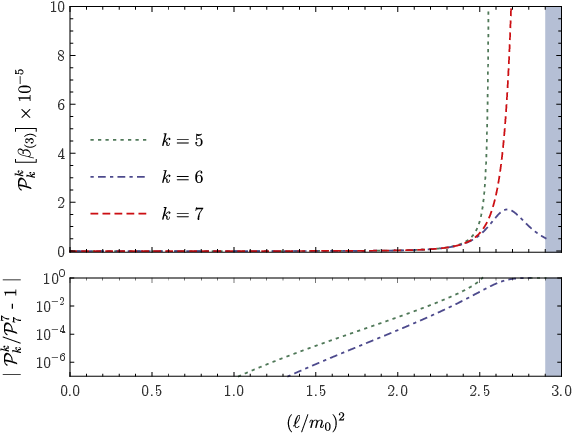}
\quad
\includegraphics[width=0.48\columnwidth]{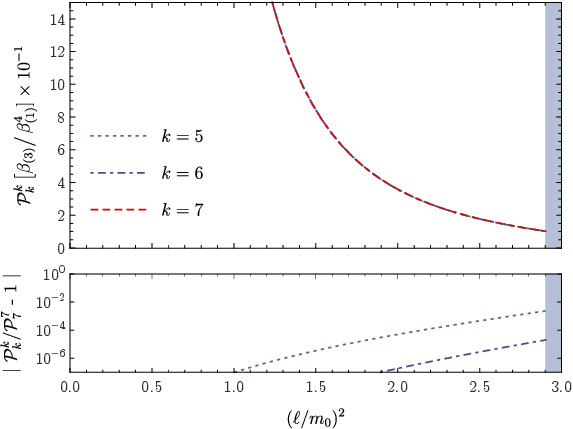}
\caption{Left panel:
Diagonal $(k,k)$-Pad\'e approximants of $\beta_{(3)}$ when $\lambda=6$.
The fractional error between the $k=6$ and $k=7$ approximants exceeds $10^{-1}$ for $(\ell/m_0)^2\gtrsim 2.50$.
Right panel:
Diagonal $(k,k)$-Pad\'e approximants of $\beta_{(3)}/\beta_{(1)}^4$ with $\lambda=6$.
The fractional error between the $k=6$ and $k=7$ approximants is smaller than $2.01 \times 10^{-5}$ in the entire $(\ell/m_0)^2$ range.
At the threshold~\eqref{eq:spontScalThreshold}, $\mathcal P^7_7[\beta_{(3)}/\beta_{(1)}^4]=10.24$ is finite.
}%
\label{fig:betasApp}%
\end{figure*}

By contrast, the approximants $\mathcal P^k_k[\beta_{(3)}]$ are not satisfactory, as shown by the left panel of Fig.~\ref{fig:betasApp} on the example $\lambda=6$.
Instead, we choose the approximants $\mathcal P^k_k[\beta_{(3)}/\beta_{(1)}^4]$, as they feature remarkable convergence, cf. the right panel of Fig.~\ref{fig:betasApp}.
We thus use:
\begin{align}
    \beta_{(3)}^{\rm Pad\acute{e}}&=\mathcal P^7_7[\beta_{(3)}/\beta_{(1)}^4]\times [\beta_{(1)}^{\rm Pad\acute{e}}]^4\,,
    \label{eq:beta3PadeDef}
\end{align}
which is also in very good agreement with our numerical results, cf. Fig.~\ref{fig:betas}.
Note that by construction, $\mathcal P^7_7[\beta_{(3)}/\beta_{(1)}^4]$ has a simple pole at $(\ell/m_0)^2=0$, but it is irrelevant and eliminated by the last factor in~\eqref{eq:beta3PadeDef}.
More interestingly, $\mathcal P^7_7[\beta_{(3)}/\beta_{(1)}^4]$ is finite at the threshold~\eqref{eq:spontScalThreshold}.
This has two consequences:
first, $\beta_{(3)}^{\rm Pad\acute{e}}$ has a pole of order four there;
and second, the quantity $\Gamma_{A}$ with $A=B$ below~\eqref{eq:GammaA} is also finite there, consistently with the results of the first column in Fig.~\ref{fig:gammaValues}.

\subsection{Numerical methods\label{App:numericalMethods}}

For our numerical calculations with {\sc Mathematica}, we apply the methods already described in Ref.~\cite{Julie:2022huo} (cf. also Appendix B there) to the theory \eqref{eq:Z2theory} truncated at $\mathcal O(\varphi^4)$, with the following changes.

We set both {\sc PrecisionGoal} and {\sc AccuracyGoal} to 17.
The integrations of Eqs.~\eqref{eq:fieldEquations} are performed in the domain $1 + 10^{-7}\le r_*\le 1.1\times 10^{10}$, with $r_*=r/r_{\rm H}$.
The quantities $\bar\varphi$ and $\alpha=-(d/m)$ defined in~\eqref{eq:NandBarVarphi} and \eqref{eq:spontScalThreshold} are then calculated at $r_* = 10^{10}$ from $\varphi\simeq \bar\varphi$ and from the ratio of
$d/r_{\rm H} \simeq -r_*^2 (d\varphi/dr_*)$ and of $m/r_{\rm H} \simeq (r_*^2/2) (dN/dr_*)$, since higher-order corrections in $1/r_*$ are negligible.

For each $(\ell/m_0)^2$ value, we derive $\alpha$ on the small-$\bar\varphi$ range
$|\bar\varphi|\le\bar\varphi_{\rm max}$, such that $|\beta_{(1)}^{\rm Pad\acute{e}}|\bar\varphi_{\rm max}=\frac{1}{6}\beta_{(3)}^{\rm Pad\acute{e}}\bar\varphi^3_{\rm max}$.
Indeed, when $|\bar\varphi|\simeq\bar\varphi_{\rm max}$, the linear and cubic contributions to~\eqref{eq:alphaExpansion} become comparable.
Note that in this range, we find that the horizon regularity condition (12) in Ref.~\cite{Julie:2022huo} is always satisfied.
We skim through $\bar\varphi$ values with increment $\Delta\bar\varphi=2\bar\varphi_{\rm max}/500$.
Finally, we define interpolations $\alpha(\bar\varphi)$ by means of B\'{e}zier curves of degree 500 using {\sc Mathematica}'s {\sc BezierFunction}, and compute the sensitivities $\beta_{(1)}=(\partial\alpha/\partial\bar\varphi)_0$ and $\beta_{(3)}=(\partial^3\alpha/\partial\bar\varphi^3)_0$ at $\bar\varphi=0$.

\subsection{Explicit expression of $\mathcal C(\hat r)$}

The function of $\hat r$ entering Eq.~\eqref{eq:phiA} reads:
\begin{align}
\mathcal C(\hat r)&=\frac{1}{2}\beta^{A}_{(1)}+\frac{1}{6}\frac{\beta^A_{(3)}}{\beta^A_{(1)}}+\left(\frac{1}{2}\beta^{B}_{(1)}+\frac{1}{6}\frac{\beta^B_{(3)}}{\beta^B_{(1)}}\right)
\left(\frac{q}{1+q}\frac{\beta_{(1)}^{A}}{\hat r}\right)^2\,.
\end{align}

\subsection{Evaluation of $\Gamma_A$ and $\Gamma_B$}

\begin{figure*}[h!]
\includegraphics[width=\columnwidth]{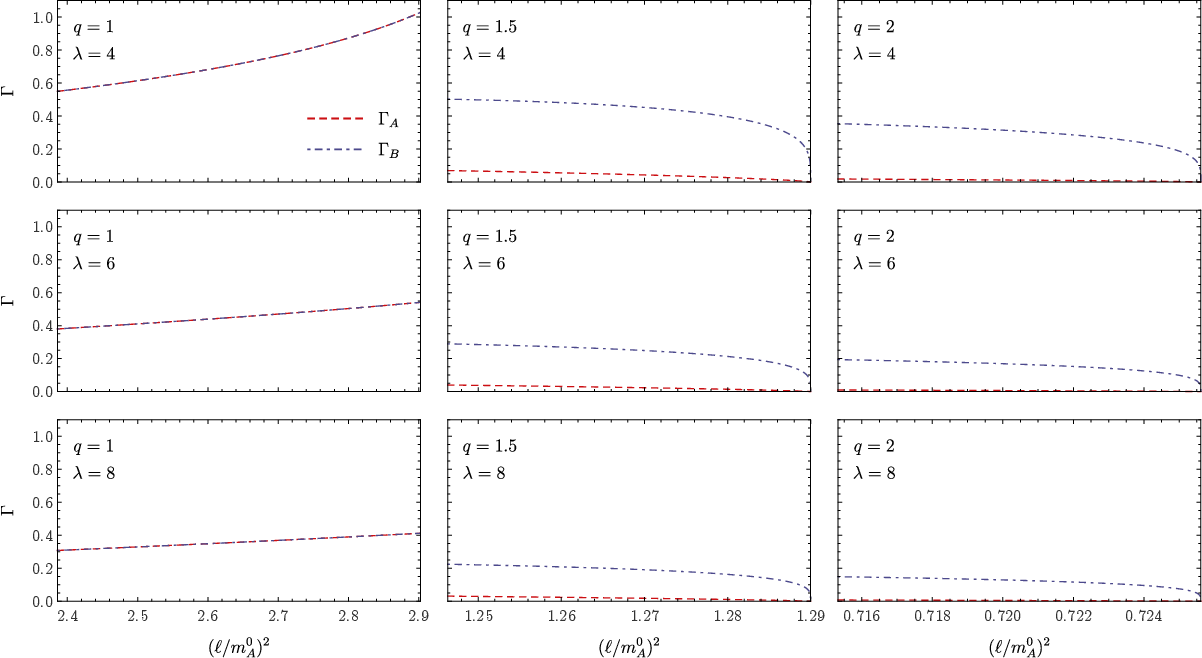}
\caption{Coefficients $\Gamma_A$ and $\Gamma_B$ for $\lambda=4,6,8$.
We set $q=1,1.5,2$, and investigate
the range
$2.39\lesssim (\ell/m_A^0)^2\lesssim 2.90$, $1.246\lesssim (\ell/m_A^0)^2\lesssim 1.290$, and $0.7153\lesssim (\ell/m_A^0)^2\lesssim 0.7256$
respectively.
The lower bounds correspond to $\hat r_{\rm DS}=3$, and at the upper bounds, $(\ell/m_B^0)^2$ reaches the spontaneous scalarization threshold~\eqref{eq:spontScalThreshold}.
}%
\label{fig:gammaValues}%
\end{figure*}

For completeness, Fig.~\ref{fig:gammaValues} evaluates $\Gamma_A$ and $\Gamma_B$ when $q=1,1.5,2$, and $\lambda=4,6,8$.
We show the full range of $(\ell/m_A^0)^2$ values relevant to DS, which is determined by the intersection of the light-shaded region and of the fixed-$q$ lines in the left panel of Fig.~\ref{fig:DsOnsetAndGrowth}.
We notice that increasing $\lambda$ or $q$ makes $\Gamma_A$ and $\Gamma_B$ smaller.
As expected, $\Gamma_A<\Gamma_B$ when $q=m_A^0/m_B^0>1$, because then the sensitivities of BH $A$ are smaller than those of BH $B$. 
Finally, observe that when $(\ell/m_A^0)^2$ reaches the spontaneous scalarization threshold~\eqref{eq:spontScalThreshold}, both $\Gamma_A$ and $\Gamma_B$ vanish, unless $q=1$.
This is simply understood when recalling that $\beta_{(1)}^{\rm Pad\acute{e}}$ and $\beta_{(3)}^{\rm Pad\acute{e}}$ have poles of order one and four there.
Thus when $q>1$, Eq.~\eqref{eq:GammaA} vanishes at the threshold.
When $q=1$, the BHs are identical, and $\Gamma_A$ given below~\eqref{eq:GammaA} remains finite instead, as explained above in the Supplemental Material.

\subsection{Nonperturbative results with a scalar background $\varphi_0$}

Eq.~\eqref{eq:phiA} shows that a BH binary can continuously branch off from the Schwarzschild configuration accross $\hat r=\hat r_{\rm DS}$, provided that we choose among two equally energetically favorable DS branches.
Here, we complete the picture in two ways:
(i) We address the effect of adding a homogeneous, nonzero scalar field background $\varphi_0$ on our BH binary.
It plays the role of a scalar field perturbation breaking the $\mathbb{Z}_2$ symmetry of the problem;
(ii) We solve for the scalar charge growth numerically, beyond the small-$\varphi$ assumption made in the main text.

\begin{figure*}[h!]
\includegraphics[width=0.48\columnwidth]{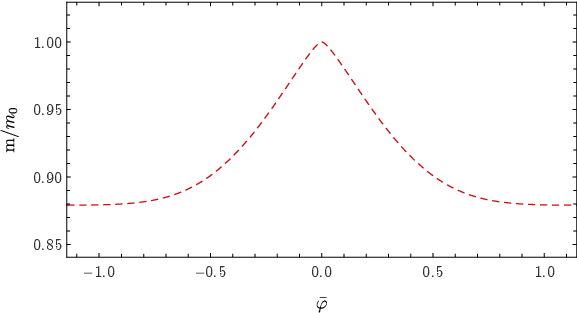}
\quad
\includegraphics[width=0.48\columnwidth]{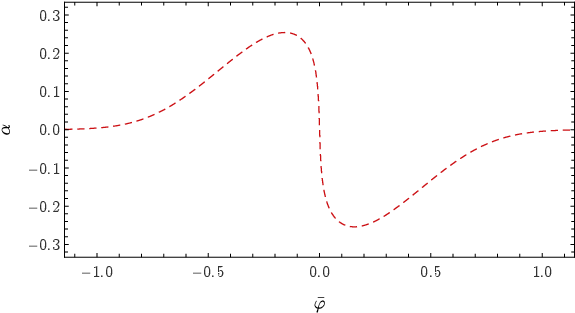}
\caption{Numerical sequence of BH solutions with fixed $\mathscr S_{\rm w}$ in the Gaussian theory with $\lambda=6$, cf.~\cite{Julie:2022huo}.
We set $(\ell/m_0)^2=2.73$ and show $m/m_0$ (left panel) and $\alpha$ (right panel) against a nonperturbative scalar background $\bar\varphi$.
}%
\label{fig:malphaNum}%
\end{figure*}

To do so, we consider the example $f(\varphi)=\frac{1}{2\lambda}[1-\exp(-\lambda\varphi^2)]$ with $\lambda=6$.
In this specific theory, Ref.~\cite{Julie:2022huo} numerically computed $m/m_0$ and $\alpha$ (but not $\beta_{(1)}$ and $\beta_{(3)}$) for a sequence of BH solutions with fixed Wald entropy $\mathscr S_{\rm w}$ and nonperturbative scalar background $\bar \varphi$.
The results are recalled in the left and right panels of Fig.~\ref{fig:malphaNum} on the example $(\ell/m_0)^2=2.73$, which is below the threshold~\eqref{eq:spontScalThreshold}.
As expected from the main text, we recover a Schwarzschild BH at $\bar\varphi=0$.
There, $m=m_0$ takes its largest value, and $\alpha=0$.
However, the spacetime also reduces to Schwarzschild in the limit $|\bar\varphi|\to\infty$, such that $\alpha\to 0$ and $m/m_0\to [1-\frac{1}{12}(\ell/m_0)^2]^{1/2}\simeq 0.879$.
The latter limit follows from evaluating the Wald entropy~\eqref{eq:WaldEntropy} at $\bar\varphi=0$ and at $|\bar\varphi|\to\infty$, and from equating the results.

\begin{figure*}[h!]
\includegraphics[width=0.48\columnwidth]{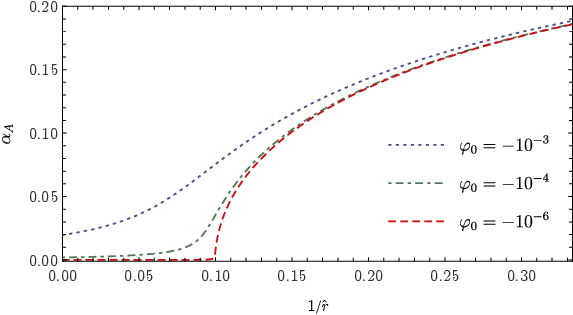}
\quad
\includegraphics[width=0.48\columnwidth]{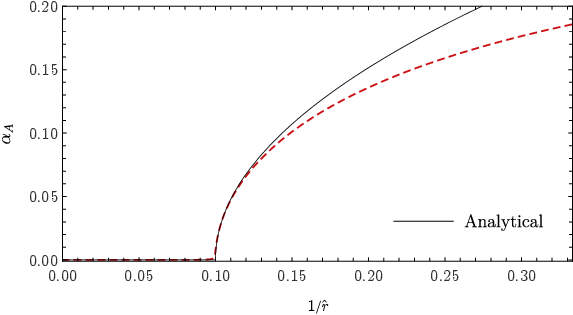}
\caption{Left panel: $\alpha_A$ against $1/\hat r$, solving~\eqref{eq:barPhiNum} nonperturbatively for various scalar backgrounds $\varphi_0$.
We consider the Gaussian theory with $\lambda=6$, and set $(\ell/m_A^0)^2=2.73$ and $q=1$.
Right panel: Comparison with the analytical, small-$\varphi$ results of the main text, cf. also the right panel of Fig.~\ref{fig:DsOnsetAndGrowth}.
}%
\label{fig:DsNum}%
\end{figure*}

We now turn to a BH binary with $(\ell/m_A^0)^2=2.73$ as in Fig.~\ref{fig:malphaNum}, and focus on the symmetric case $A=B$ for simplicity.
In presence of a scalar background $\varphi_0$, the system \eqref{eq:CoulombBackgrounds} is replaced by a single equation:
\begin{align}
    \bar\varphi_A=\varphi_0-\frac{m_A(\bar\varphi_A)\alpha_A(\bar\varphi_A)}{r}\,.\label{eq:barPhiNum}
\end{align}
Given $\varphi_0$ and $\hat r=r/2m_A^0$, we can solve numerically for $\bar\varphi_A$ above, using the results of Fig.~\ref{fig:malphaNum} (see e.g.~\cite{Palenzuela:2013hsa} for the example of NSs with fixed baryonic number in ST theories).
The resulting $\alpha_A$ is shown in Fig.~\ref{fig:DsNum} as a function of $1/\hat r$ when $\varphi_0=-10^{-3},-10^{-4},-10^{-6}$.
For large $\hat r$ values, the solution to~\eqref{eq:barPhiNum} is unique, and since it is negative $\bar\varphi_A<0$, we have $\alpha_A>0$.
For sufficiently small $\hat r$ values (always verifying $\hat r<\hat r_{\rm DS}$), up to two more solutions $\bar\varphi_A>0$ to~\eqref{eq:barPhiNum} appear, but we discard them.
Indeed, they have a smaller magnitude, and are thus energetically disfavored since they correspond to BHs with larger ADM masses, cf. the left panel of Fig.~\ref{fig:malphaNum}.
Moreover, they are disjoint from the first branch, because they have the opposite sign.
Thus, the system cannot leave the first branch without losing adiabaticity.
As shown by the left panel, the scalar hair grows with $1/\hat r$, and as $|\varphi_0|$ is decreased, an abrupt transition forms close to $\hat r\simeq 10$.
Note that at the light-ring $\hat r_{\rm LR}=3$ and for $\varphi_0=-10^{-6}$, we find $\alpha_A\simeq 0.186$.
The right panel shows that the agreement between the analytical results of the main text (based on the expansion~\eqref{eq:Z2theory}) and our nonperturbative analysis here is excellent, as expected, close to the DS threshold.
We recall that $\hat r_{\rm DS}=10$ according to the left panel of Fig.~\ref{fig:DsOnsetAndGrowth} when $(\ell/m_0)^2=2.73$ and $q=1$.

The case $\varphi_0>0$ is straightforwardly inferred from the discussion above, given the $\mathbb{Z}_2$ symmetry of the theory.

\end{document}